\documentclass[utf8]{frontiersFPHY} 

\setcitestyle{square} 
\usepackage{url,hyperref,lineno,microtype,subcaption}
\usepackage[onehalfspacing]{setspace}
\usepackage{color}

\definecolor{myred}{RGB}{220, 20, 60}

\newcommand{\pythia}{{\ttfamily PYTHIA} }
\newcommand{\herwig}{{\ttfamily HERWIG} }

\newcommand{\lsim}   {\mathrel{\mathop{\kern 0pt \rlap
  {\raise.2ex\hbox{$<$}}}
  \lower.9ex\hbox{\kern-.190em $\sim$}}}
\newcommand{\gsim}   {\mathrel{\mathop{\kern 0pt \rlap
  {\raise.2ex\hbox{$>$}}}
  \lower.9ex\hbox{\kern-.190em $\sim$}}}

\def\keyFont{\fontsize{8}{11}\helveticabold }
\def\firstAuthorLast{V. Gammaldi} 
\def\Authors{Viviana Gammaldi\,$^{1}$
}
\def\Address{$^{1}$Instituto de F\'isica Te\'orica (IFT), Universidad Autonoma de Madrid, Madrid, Spain. 
}
\def\corrAuthor{Corresponding Author}

\def\corrEmail{viviana.gammaldi@uam.es}

\begin{document}
\onecolumn
\firstpage{1}

\title[TeV Dark Matter]{Multimessenger TeV Dark Matter:\\ a mini review} 

\author[\firstAuthorLast ]{\Authors} 
\address{} 
\correspondance{} 

\extraAuth{}

\maketitle

\begin{abstract}

We briefly review the general insight into the indirect searches of dark matter. We discuss the primary equation in a three-level multimessenger approach (gamma rays, neutrinos and antiprotons), and we introduce the reader to the main topics and related uncertainties (e.g. dark matter density distribution, cosmic rays, particle physics). As an application of the general concept, we focus on the multi-TeV dark matter candidate among other weak interactive massive particles. We present the state-of-the-art on this sub-field, and we discuss open questions and experimental limitations. 

\tiny
 \keyFont{ \section{Keywords:} dark matter, cosmic rays, TeV energy scale, indirect searches, simulations, density distribution} 
\end{abstract}

\section{Introduction: dark matter, an open question }
More than 80 years ago, F. Zwicky applied the virial theorem to the Coma Cluster and determined that a large amount of non-luminous matter must be present to keep the system bound together \cite{Zwicky}; nearly 40 years later, V. Rubin observed similar gravitational evidences studying the rotation curve of spiral galaxies \cite{Rot_curve1, Rot_curve2, Rot_curve3, Rot_curve4}. From then on, many astrophysical and cosmological evidences hint to some inconsistency in our understanding of the Universe as a whole. Many theories have been proposed in order to account for the gravitational observations: they include both modified gravity (e. g. \cite{Milgrom}) or a dark component of matter \cite{lensing1, lensing2, Planck1, Planck2}. In particular, the need for non-baryonic Dark Matter (DM) is favored by a variety of independent estimates of the matter density in the Universe, that points to a value larger than the value provided by baryons alone, according to nucleosynthesis (see e.g. \cite{Bergstrom:2000pn} for a general overview).   DM candidates cover a broad range of masses from $10^{-35}$ to $10^{18}$ GeV \cite{DMcan} (see e.g. \cite{Bertone:2004pz}). Among them, and beyond the Standard Model (SM) of particle physics, the Weak Interactive Massive Particle (WIMP) represents one plausible candidate, compatible with both cosmological constraints and large scale structure (galaxies and galaxy clusters) 
formation and evolution models and simulations \cite{Naab}. In particular, TeV WIMP stands as an open possibility and one of the next frontiers for the DM community \cite{TeVNature}. Up to masses of 100 TeV, DM candidates still conserve cosmological properties of thermal candidates. Thermal relics were as abundant as photons in the primordial hot plasma, being freely created and destructed in pairs in the thermal bath. Due to the cooling of the Universe, their relative number density started then being suppressed as annihilations proceeded. When the temperature dropped below their mass, the annihilation processes froze out and their final abundance would be the observed $27\%$ of the whole content of the Universe nowadays \cite{Planck1,Planck2}.\\
\\
 WIMPs searches based on different approaches and methodologies, have been developed in order to investigate different energy scales. They are commonly classified in three main classes: experiments of direct and indirect searches of DM and colliders. DM searches at colliders, such as the Large Hadronic Collider (LHC) among others \cite{Coll, DMparticle, Munoz}, focus on the possibility to produce DM particles by the interaction of two SM particles and the subsequent production of unknown particles (SM-SM $\rightarrow$ DM-DM, see e.g. Fig. 2.1 of \cite{VGthesis} for a schematic visualization of different processes). 
Due to both experimental and theoretical limitations, the higher particle mass that can be studied at this experiment is few TeV, and they are strongly dependent on the particle physics model of interest \cite{LHC}. Therefore, the study of particle physics nature of multi-TeV DM candidate at colliders is a challenge and represents a new frontier in physics. Similar limitations affect the experiments of direct searches \cite{PhysRevD.31.3059, PhysRevD.30.2295, Baudis:2014naa, DMparticle}. The latter are underground experiments designed in order to investigate the SM-DM $\rightarrow$ SM-DM interaction, that is, the scattering angle between the prospective DM particle within the Milky Way halo with heavy nuclei. This kind of experiment mainly addresses DM particle mass of $1-10^4$ GeV 
depending on the spin dependence \cite{direct, DMparticle, Munoz}. The DAMA Collaboration claimed for a periodic signal that could be explaind with a DM particle mass of fews (tens) GeV \cite{Bernabei:2008yi,Bernabei:2013xsa,Bernabei:2018yyw,Baum:2018ekm}. However, strong tension emerges between the DAMA/NaI and DAMA/LIBRA claim and the null results from several underground experiments \cite{Savage:2008er}, such as CDMS \cite{Ahmed:2008eu}, XENON10 \cite{Angle:2008we}, CRESST I \cite{Lang:2009ge}, CoGeNT \cite{Aalseth:2008rx}, TEXONO \cite{Lin:2007ka}, and Super-Kamiokande (SuperK)\cite{Desai:2004pq}. \\
\\
Searches of multi-TeV DM candidates can be addressed by means of cosmic-ray experiments, that allow to investigate the energy range from few MeV to PeV. In particular, detectors of very high energy (VHE) cosmic rays investigate the TeV energy scale. Indirect searches of DM focus on the DM-DM $\rightarrow$ SM-SM interaction, that is the production of SM particles by DM annihilation or decay events in astrophysical targets \cite{VGthesis}, with a process similar to that happening in the primordial plasma before particles decoupling.
The benchmark thermal annihilation cross-section is $\langle \sigma v\rangle_\text{ann}=3\times10^{-26}\text{cm}^3\text{s}^{-1}$ and the decay half-life is tuned to $\tau_\text{dec}\approx 10^{26} s$. WIMPs annihilate or decay into SM particles, which then produce secondary fluxes of cosmic rays (gamma-rays, neutrinos, antimatter) that are collected by detectors. This class of searches are independent of the particle physics model, and only depends on the energy of the primary annihilation/decay event. The multimessanger approach for DM searches implies to collect the complementary information given by different cosmic rays and experiments. \\
\\
In this review, we focus on a subclass of cosmic-ray experiments. We briefly introduce the reader to the fundamentals of multimessenger approach for indirect search of WIMPs (Section \ref{multimessenger}). In particular, in Section \ref{multiTeV} we will discuss recent results of TeV DM studies. We provide very general information about brane world theory as a possibility for multi-TeV DM candidates in Section \ref{TeVnature}. Finally, we will tray the main conclusions and prospectives for future studies in Section \ref{conclusions}.

\section{Multimessenger approach to indirect searches of DM.}
\label{multimessenger}

The multimessenger approach is the next frontier for DM searches. Signatures of DM annihilation or decay events in astrophysical sources may be observed in cosmic-ray fluxes by Cerenkov telescopes such as VERITAS \cite{VERITAS}, HESS \cite{HESS}, MAGIC\cite{MAGIC}, HAWC \cite{HAWC} and CTA \cite{CTA}; neutrino telescopes such as ANTARES \cite{ANTARES} or IceCube \cite{IceCube}; satellites such as PAMELA \cite{PAMELA}, AMS \cite{AMS02} and Fermi \cite{Fermi} or ballon experiments like CAPRICE \cite{CAPRICE} or BESS \cite{BESS}( for a general overview see e.g. \cite{indirect}).  
The secondary products of annihilation and decay of DM particles
contribute to the cosmic-ray differential flux as \cite{Gammaldi:2014noa}:

\begin{equation}
\frac{d \Phi_{\text{cr-DM}}}{dE}=\eta_{\text{cr}}\cdot\sum^2_{a=1} \sum^{\text{ SM channels} }_i \frac{\zeta^{(a)}_i}{a}
\frac{dN^{(\text{cr})}_{i}}{dE}
 \cdot \frac{\kappa^{(a)}_{\text{cr}}}{4 \pi M^a},
 \label{phigen}
\end{equation}

where:

\begin{itemize}
\item $\eta_\text{cr}$ depends on the secondary particles of interest (cosmic rays) and their propagation; $\eta_\text{cr}=1$ for gamma rays, otherwise it depends on neutrino oscillations or the velocity of the charged particle for antimatter studies; 
\item the total flux is given by decay  ($a=1$) and/or annihilation ($a=2$) events of DM particle into the $i$-th SM particle (annihilation/decay channel). The $\zeta$ factor discerns between these two cases: $\zeta^{(1)}_i=1/\tau_i^{\text{decay}}$ and $\zeta^{(2)}_i=\langle\sigma_i v\rangle$ are respectively the inverse of the decay time and and thermal averaged annihilation cross section times velocity. The probability that DM annihilates or decays into the $i$-th channel depends on the nature of DM;
 
\item the differential number $dN_i^{(\text{cr})}/dE$ of cosmic rays produced at the source by subsequent events of annihilation or decay of SM particles is simulated by means of Monte Carlo events generator software, such as \pythia or \herwig.  A Particle Physicist Cookbook for Dark Matter (PPPC4DM) \cite{Cirelli} provides the cosmic-ray fluxes for an immediate application. In particular, electroweak corrections are important for multi-TeV events \cite{EW}. Some uncertainties may be introduced in the evaluation of both the $\zeta^{(a)}_i$ and $\kappa_\text{cr}$ factor due to the choice of the Fortran or C++ versions of \pythia or \herwig software, as discussed in \cite{MC} for gamma-ray fluxes;
\item the $\kappa_\text{cr}$ factor depends on the astrophysics of DM distribution as well as on the cosmic-ray propagation.
For \textit{neutral} cosmic-rays (n-cr) (e.g. gamma rays and neutrinos) it is the so-called astrophysical J-factor 

\begin{equation}
\kappa^{(a)}_{\text{n-cr}}\equiv{\langle J \rangle}_{\Delta\Omega}= \frac{1}{\Delta\Omega}\int_{\Delta\Omega}\text{d}\Omega\int_{l_\text{min}}^{l_\text{max}} \rho^{(a)} [r(l)] dl(\alpha)\,.
\end{equation}

Here, $\rho(r)$ is the DM density distribution. The line of sight (l.o.s) $l$ is the distance from the observer to any observed source. The radial distance $r$ from the center of the target to a given point inside it, is related to $l$ by $r^2 = l^2 + d^2 -2d l \cos \alpha$, where $d$ is the distance from the Earth
to the center of the target. The distance from the Earth to the edge of the DM halo in the direction $\alpha$ is
$l_\text{min/max}(r,d,\alpha) = d \cos \alpha + \sqrt{r^2-d^2 \sin \alpha}$. For \textit{neutral} particles,  \textit{directional} observations are achievable. In this case, the flux must be averaged over the solid angle of the detector, that is typically of order of $\Delta \Omega = 2 \pi ( 1 - \cos \theta )$, being $\theta$ the angular resolution of the telescope. Further details about the calculation of the astrophysical factor can be found e.g. in Appendix B of \cite{Gammaldi:2017mio}. \\
For \textit{charged} cosmic-rays (c-cr), \textit{directional} observations are \textit{not} feasible. In fact, charged particles observed in a given direction might have been produced everywhere in the sky. In this case the $\kappa_\text{c-cr}$ factor is proportional to a diffusion term
\begin{equation}
\kappa_\text{c-cr}^{(a)}\equiv \left(\frac{\rho_\odot}{M}\right)^{(a)}R_{\text{c-cr}}(r_\odot, E).
\label{kappa}
\end{equation}

The diffusion factor at the position of the Sun $R_{\text{c-cr}}(r_\odot, E)$ for \textit{charged} cosmic rays ($e^\pm, p^\pm$) is the solution of a diffusion equation that depends on the particle of interest and DM distribution. It describes the diffusion of particles in the Galaxy and the production of secondary cosmic rays due to the interaction with the Interstellar Medium (ISM). The final flux at the position of the Earth also includes Solar magnetic field effect \cite{Perko}.

\end{itemize}

Each of the previous points require further investigations,  yet this is beyond the scope of this review. In the following we will apply such a primary equation to the specific case of TeV DM candidates. We will show recent results for several cosmic rays (cr = gamma rays, neutrinos and antiprotons), focusing on annihilation events ($a=2$ and $\zeta^{(2)}_i=\langle\sigma_i v\rangle$).

\section{Indirect searches and multi-TeV DM}
\label{multiTeV}

\begin{figure}[h!]
\begin{center}
\includegraphics[width=14cm]{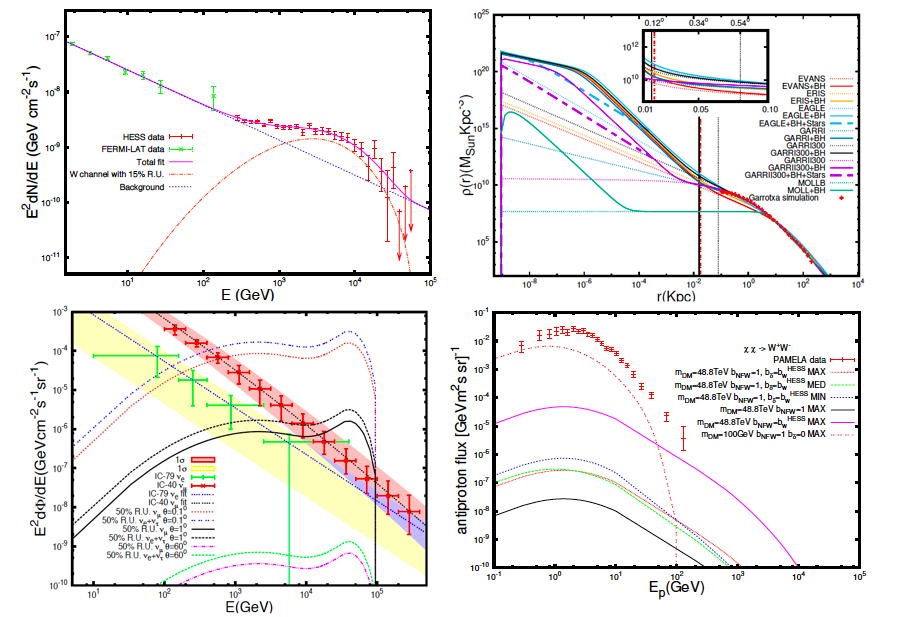}
\end{center}
\caption{\footnotesize{ Upper Left: Figure from \cite{ThesisVG}. The combination of the gamma-ray flux expected by DM annihilation events with a power-law background component well fits the gamma-ray spectra of Fermi-LAT and HESS data with a thermal DM particle of mass of $48.8$ TeV that annihilates in the $W^+W^-$ boson channel \cite{Cembranos:2013fya, Cembranos:2012nj}. Upper Right: Figure from \cite{Gammaldi:2016uhg}. The enhancement of $10^3$ required in order to fit the gamma-ray spectra can be explained by the J-factor of a BH-induced DM-spike in hydrodinamical N-body simulations with a cusp profile. Lower Left: Figure from  \cite{neutrinos}. The expected neutrino flux from $48.8$ TeV DM candidate is compared with the atmospheric neutrino background detected by IceCube. Lower Right: Figure from \cite{Cembranos:2014wza}. the expected antiproton flux from $48.8$ TeV DM candidate is compared with the PAMELA data. The total antiproton flux takes into account the diffusion of antiprotons produced by \textcolor{red}{multi-}TeV DM in the halo and the extra component given by the DM-spike at the GC.}}\label{fig:1}
\end{figure}

 Multi-TeV DM candidates have been proposed in literature e.g. in order to explain the cut-off at TeV energy scale observed by the HESS telescopes at the Galactic Center (CG) \cite{HESSGCdata, HESS10y}. The interpretation of these fluxes as DM signal has been widely discussed from the very early days of the publications of the observed data \cite{PhysRevLett.97.221102, Bergstrom:2004cy, PhysRevLett.95.241301, TeVProfumo}. At a first moment, it was concluded that the spectral features of these gamma rays disfavoured the DM origin \cite{PhysRevLett.97.221102, 2005ApJ...619..306A}. More recently, combined analyses of Fermi LAT and HESS data have allowed new interpretations \cite{Cembranos:2013fya, Cembranos:2012nj, Zaharijas}.\\
\\
 Before going into further details and in order to avoid any misunderstanding, it should be noted that we do not refer to Fermi-LAT signal commonly known as the GeV-excess. In fact, Fermi LAT among other experiments (e.g. MAGIC \cite{Segue1Magic} ), have allowed to set stringent constraints on the DM particle mass and annihilation cross-section \cite{DiMauro:2015tfa, Lisanti:2017qoz, Ahnen:2016qkx, Fermi-LAT:2016uux, SMC, LindenExt1} as well as to increase the number of claims of gamma-ray signatures from DM \cite{LindenExt, RetII, Fermi-LAT:2016uux, PhysRevD.93.043518}. However, most of these studies deal with DM particle mass between few MeV and hundreds of GeV. In particular, the GC-excess has been largely interpreted both in terms of DM signatures \cite{Calore1, Calore2, Bertone:2015tza, Caron:2015wda, Cerdeno:2014cda, Huang:2014cla, Alves:2014yha, Karwin:2016tsw} and of astrophysical signal from Millisecond Pulsars (MSPs)\cite{Cholis:2014lta, Petrovic:2014xra, Brandt:2015ula, Lee:2015fea, Hooper:2015jlu, Bartels:2015aea, Hooper:2016rap, Haggard:2017lyq, Hooper:2018fih}. On the other hand, the combined analysis of Fermi-LAT data with TeV experiments open new avenues to different interpretations. If the high energy (HE) Fermi-LAT data are assumed as the background component for the HESS observations in the GC region, the fit of the VHE HESS data improves with respect to previuos works \cite{Cembranos:2013fya, Cembranos:2012nj, VGthesis, Zaharijas}. In particular, the HESS data show a spectral cut-off within a region of tens of parsecs from the GC \cite{HESS10pc} and no spectral cut-off in a region of hundreds of parsecs from the GC. The latter emission let researchers think about the existence of a pevatron accelerator\cite{HESSPeV}, and an astrophysical origin from GeV to TeV energy scale \cite{Gaggero:2017jts}. Some efforts have been also pursued to explain the TeV cut-off as a combined signal of TeV DM and MSPs \cite{Lacroix:2016qag}. 
\\
\\
In the following, we assume the astrophysical origin for the GeV-excess, and we focus on the multi-TeV DM interpretation of the HESS data. In the upper left panel of Fig.\ref{fig:1}, we show the fit to gamma-ray spectra of Fermi-LAT and HESS data with a thermal DM particle mass of $48.8$ TeV that annihilates in the $W^+W^-$ boson channel \cite{Cembranos:2013fya, Cembranos:2012nj}. Similar fits can be found in \cite{Zaharijas}. This multi-TeV DM candidate is still a subject matter of discussion. In fact, the TeV DM hypothesis in this region needs an enhancement factor that, in the case of thermal DM particle, could be explained as increasing DM density throughout the GC with respect to the benchmark NFW profile \cite{Cembranos:2013fya, Cembranos:2012nj}. Hydrodynamic simulations together with a Black Hole (BH) induced DM-spike may explain the required enhancement of $10^3$ in the J-factor \cite{Gammaldi:2016uhg}, and the radial dimension of the DM-spike depends on the initial DM profile, that is a cuspy or core \cite{Gammaldi:2016uhg, Lacroix:2018zmg, Gammaldi:2017cvj}. The DM spike that corresponds to several DM halo density profiles is shown in the upper right panel in Fig \ref{fig:1}. \\
\\
In the framework of the multimessanger approach to DM searches, further investigation has been addressed in order to study the possibility to detect a neutrino signal originated by the same multi-TeV DM candidate, that may explain the observed cut-off in gamma rays. For neutrino searches, in Eq. \ref{phigen} $\eta_\nu=\text{P}_{fp}$ where $\text{P}_{fp}$ are the elements of the symmetric $3\times3$ matrix which takes into account the neutrino oscillation effects from the neutrino flavor ($\nu_p$) produced at the source and the neutrino flavor ($\nu_f$) observed at the telescope. The astrophysical J-factor is the same as for gamma rays (for more details see \cite{neutrinos}). In order to get a $2\sigma-5\sigma$ neutrino signature from the DM candidate of $\approx50$ TeV annihilating in $W^+W^-$ SM channel (with a minimum of 2-years of exposition time and $50\text{m}^2$ of detector effective area) the IceCube telescope should have resolution angle $\theta\lsim 0.72^\circ$ and low energy threshold $\approx 1-2$ TeV.  Currently, the IceCube resolution angle is worst than $5^\circ$, making unrealistic this kind of required observation \cite{Aartsen:2017ulx}. This is shown in the lower left panel of Fig. \ref{fig:1}. \\
Moreover, the same primary DM annihilation event, if constituting the origin of the observed cut-off in TeV gamma rays, may also produce leptonic or hadronic counterparts. The production of a concrete particle will induce 
secondary production that would affect mainly the diffuse signal through hadronic emission by inelastic proton collision with the interstellar gas, inverse Compton scattering of interstellar radiation by cosmic-ray electrons and positrons, or Bremsstrahlung. The $e^\pm$ and $p \bar p$ data from ATIC/PPB-BETS, PAMELA, Fermi LAT and AMS have been largely studied \cite{Cirelli:2012tf}. 
On the one hand, low energy data are consistent with astrophysical primary sources \cite{Serpicoreview, Berezhko:2014yda, 0004-637X-845-2-107,Giesen:2015ufa, DiMauro:2014iia} (yet see e. g. \cite{Belotsky:2016tja, Belotsky:2014haa,Giesen:2015ufa} for different interpretation as DM). 
In this sense, antiproton data can be used to characterize diffusion models of charged particles along the Galaxy or to constrain new physics, whose antiproton flux may be identified upon the diffusion background. As discussed in the introduction, the $\kappa$-factor is given by Eq. (\ref{kappa}) for charged cosmic rays; instead $\eta_\text{p} \propto v_\text{p}$, where $v_\text{p}$ is the antiproton velocity (see \cite{Cembranos:2014wza} for more details). The antiproton flux generated by $\approx 50$ TeV DM candidate distributed in the halo with a possible enhancement at the GC appears to remain below the antiproton flux measured by PAMELA (lower right panel in Fig. \ref{fig:1}). Increasing the maximum energy threshold in antiproton data would allow to detect some signature on the extrapolated background. \\
On the other hand, the very recent AMS-02 positron data \cite{AMS022019} are consistent with primary emission from astrophysical sources at low energy scale \cite{Serpicoreview}, yet a finite energy cut-off at  $810^{+310}_{-180}$ GeV is established with a significance of more than $4\sigma$. Such a cut-off may be predominantly originate either from DM or from other astrophysics. We would like to invoke the possibility - that has not yet been explored -  to study these data within the multi-TeV DM framework. 
\\
\\  
Complementary to the GC region, dwarf galaxies represent excellent targets for the indirect searches of DM. The GC represents a very appealing target, due to its closeness and the big amount of DM in the region, yet the complex nature of this area makes the identification of the sources quite difficult, as we discussed so far. On the other hand, dwarf galaxies are close DM dominated structures with low astrophysical background. Well-known dwarf spheroidal (dSph) galaxies are pressure-supported systems where the contamination from intrinsic astrophysical sources is negligible \cite{Winter:2016wmy}. In fact, they host an old stellar population of low-luminosity and do not possess gas. These objects have been studied at TeV energy scale by several gamma-ray telescopes, such as VERITAS \cite{dSphVERITAS}, HESS \cite{dSphHESS, dSphHESSWIMP}, MAGIC \cite{dSphMAGIC} and HAWC \cite{dSphHAWC}. The $\approx 50$ TeV DM $\rightarrow W^+W^-$ candidate results to be compatible with all these exclusion limits. In fact, the enhancement factor required to fit the HESS spectral cut-off at the GC, has to be understood as 
due to the local environment. Indeed, it would not be applied to dwarf satellites unless any BH is detected
  \cite{Gonzalez-Morales:2014eaa}.\\
Recently, it has been shown that the astrophysical contamination in gamma rays is negligible also in rotationally-supported dwarf irregular (dIrr) galaxies \cite{Gammaldi:2017mio}. 
Because Active Galactic Nuclei are not observed in dIrr galaxies, the background is expected to be negligible at TeV scale.  
In the left panel of Fig. \ref{fig:2} we show the preliminary study of the HAWC collaboration on this new class of targets for the analysis of 1-year of data taking \cite{Cadena:2017ldx}. Although the results do not reach the thermal annihilation cross-section value and bounds are not competitive with respect to those obtained at lower WIMP mass scales, they represent new results for the TeV DM candidate by means of a previously unexplored kind of DM targets, i.e. dIrr galaxies.  


\section{The multi-TeV DM particle nature}
\label{TeVnature}

\begin{figure}[h!]
\begin{center}
\includegraphics[width=18cm]{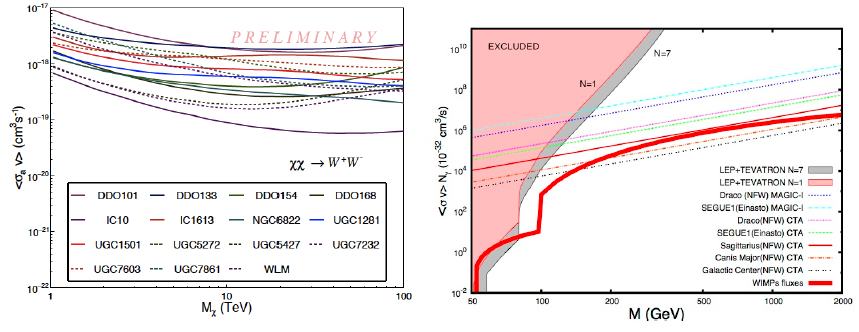}
\end{center}
\caption{\footnotesize{Left Panel: Figure from \cite{Cadena:2017ldx}. Constraints on the DM mass and annihilation cross-section by the study of several dwarf irregular galaxies at TeV energy scale, with 1-year of data of the HAWC observatory. Right Panel: Figure from \cite{Cembranos:2011hi}. Study of detectability of branon DM with Cherenkov telescopes.}}
\label{fig:2}
\end{figure}

The particle nature of prospective multi-TeV DM candidate has been investigate since the first analysis of the HESS data. Among others DM candidates \cite{DMcan} and beyond the SM, the largest neutralino masses appears unlikely to explain the HESS data \cite{TeVProfumo}. Few models could naturally produce DM particle mass from fews to tens TeV - see e.g. dark atoms \cite{Belotsky:2014haa,Belotsky:2016tja}) or minimal DM models \cite{Cirelli:2009uv,Garcia-Cely:2015dda}. Brane World Theory may naturally produce thermal DM candidate up to masses of 100 TeV \cite{BWDM}. In brief, in the framework of extra-dimensions, and in the particular case of four-dimensional effective phenomenology, massive branons are new pseudoscalar fields which can be understood as the pseudo-Goldstone bosons corresponding to the spontaneous breaking of translational invariance in the bulk space produced by the presence of the brane. They are prevented from decaying into SM particles by parity invariance on the brane. Limits on the model parameter from three-level processes in colliders are given by HERA, Tevatron, LEP-II and LHC \cite{Cembranos:2011cm}, and prospects for ILC and CLIC can be found in \cite{Achard:2004uu, CREMINELLI2001125}. As introduced before, strong experimental limitations in direct searches and colliders affect the study of branons as multi-TeV WIMP candidates. In the right panel of Fig. \ref{fig:2} we show the prospect of detectability of branons DM with the future Cherenkov telescope array (CTA), among the others \cite{Cembranos:2011hi}.

\section{Conclusions}
\label{conclusions}

We have briefly discussed general aspects of the multimessenger approach to the indirect searches of DM focusing on the TeV energy scale. Multi-TeV DM candidate has started to being considered after the advent of the last generation of gamma-ray telescopes and recent observations of the GC region. The $\approx 50$ TeV DM $\rightarrow W^+W^-$ candidate combined with a power-law background component, well fits the VHE gamma-ray spectral cut-off observed by HESS in the inner 10 parsecs at the GC combined with the HE Fermi-LAT data. Additionally, multimessanger searches via both neutrino and antimatter fluxes have been address to better investigate such a heavy DM hypothesis for the GC. DSph galaxies have been also scanned at TeV energy scale by MAGIC, HESS, VERITAS and HAWC telescopes, resulting in compatible constraints. The next generation of both experiment (such as future CTA) and previously unexplored kind of DM targets (i.e. dIrr galaxies) will improve more and more the constraints via indirect searches. Indirect searches of DM may be performed also by the study of the synchrotron radiation, that is a radio signal emitted by the interaction of the secondary fluxes of charged particles produced in DM annihilation or decay events with the magnetic field of the target. This signal may be detected by the SKA telescope, also for the particular case of branon DM candidate in the GC \cite{Bacon:2018dui, Bull:2018lat} or in dwarf galaxies (J.A. R. Cembranos, A. de la Cruz-Dombriz, V.G., M. Mendez-Isla, in preparation). Finally, on the theoretical side, multi-TeV branon DM candidate represents an appealing possibility among others. Unfortunately, strong experimental limitations makes it very difficult to set further constraints on the nature of the multi-TeV DM particle through direct searches and colliders. Therefore, the study of particle physics nature of multi-TeV DM candidate at underground laboratories is a challenge and represents a new frontier in physics.

\section*{Conflict of Interest Statement}

The authors declare that the research was conducted in the absence of any commercial or financial relationships that could be construed as a potential conflict of interest.

\section*{Author Contributions}

VG contributed conception and design of the study, performed the analysis and organized the results of this mini review. 


\section*{Funding}
This work has been supported by JUAN DE LA CIERVA-FORMACIÓN FJCI-2016-29213, by the Spanish Agencia Estatal de Investigación through the grants FPA2015-65929-P (MINECO/FEDER, UE) and IFT Centro de Excelencia
Severo Ochoa SEV-2016-0597, by INFN project QGSKY, by the Agencia Estatal de Investigaci\'on (AEI) y al Fondo Europeo de Desarrollo Regional (FEDER) FIS2016-78859-P(AEI/FEDER, UE) and partially by the H2020 CSA Twinning project No.692194 ORBI-T-WINNINGO.


\section*{Acknowledgments}
The author acknowledges the support of the Spanish Red Consolider MultiDark FPA2017-90566-REDC. The section 2 \textit{Multimessenger Approach to indirect searches of DM} was adapted with permission of the EPJ Web of Conferences 121, 06003 (2016), both the upper-left panel in Fig.1 and right panel in Fig.2  were adapted with permission of the Universidad Complutense de Madrid. The lower-right panel in the same Fig. 1 was reused with permission of  IOPscience. The license agreements between the American Physical Society ("APS") and V.G. with licence numbers: RNP/18/DEC/009954 and RNP/18/DEC/009955 respectively allows the republication of the figures in the upper-right and lower-left panels in Fig.1 of this manuscript. VG also thanks J.A.R. Cembranos, R. Lineros, C. Muñoz, E. Rold\'an and M. A. S\'anchez-Conde for useful discussions.




\bibliography{ISTeVDM_frontiers_arXiv}

\end{document}